\documentclass[aps,prb,showpacs,twocolumn,superscriptaddress]{revtex4}
\usepackage{latexsym}
\usepackage{amssymb}
\usepackage{bm}
\usepackage{graphicx}
\begin{document}
\title{Diffusive transport in graphene: the role of interband correlation}
\author{S. Y. Liu}
\email{liusy@mail.sjtu.edu.cn}
\affiliation{Department of Physics,
Shanghai Jiaotong University, 1954 Huashan Road, Shanghai 200030,
China}
\author{X. L. Lei}
\affiliation{Department of Physics, Shanghai Jiaotong University,
1954 Huashan Road, Shanghai 200030, China}
\author{ Norman J. M. Horing}
\affiliation{Department of Physics and Engineering Physics,
Stevens Institute of Technology, Hoboken, New Jersey 07030, USA}
\begin{abstract}
We present a kinetic equation approach to investigate dc transport
properties of graphene in the diffusive regime considering
long-range electron-impurity scattering. In our study, the effects
of interband correlation (or polarization) on conductivity are
taken into account. We find that the conductivity contains not
only the usual term inversely proportional to impurity density
$N_i$, but also an anomalous term that is linear in $N_i$. This
leads to a minimum in the density dependence of conductivity when
the electron density $N_{\rm e}$ is equal to a critical value,
$N_c$. For $N_{\rm e}>N_c$ the conductivity varies almost linearly
with the electron density, while it is approximately inversely
proportional to $N_{\rm e}$ when $N_{\rm e}<N_c$ in the diffusive
regime. The effects of various scattering potentials on the
conductivity minimum are also analyzed. Using typical experimental
parameters, we find that for RPA screened electron-impurity
scattering the minimum conductivity is about $5.1\,e^2/h$ when
$N_{\rm e}\approx 0.32N_i$.
\end{abstract}
\maketitle
\section{Introduction}
Recently, graphene has attracted a great deal of experimental and
theoretical interest.\cite{NV,NV1,Zhang,Zhang1,Review,Review1} In
this two dimensional system, low energy electrons behave as massless
relativistic fermions due to their linear energy spectrum around two
nodal points in the Brillouin zone.\cite{ES} Such unusual electronic
properties lead to high mobility as well as a long mean free path at
room temperature, making graphene a promising candidate for future
electronic applications.\cite{Berger}

In the first experiments involving graphene, it was found that
there exists a finite "residual" conductivity, $\sigma_{\rm res}$,
in the carrier-density dependence of conductivity at zero gate
voltage, and its value is about $4 e^2/h$. Furthermore, it was
found that the conductivity varies linearly with carrier density
when it is large. Much theoretical effort has been devoted to
quantitatively explain
 the observed "residual" longitudinal conductivity. Actually, the
 existence of such a "residual" conductivity in perfect
 (scattering-free) single-layer graphene
  was predicted long before its
 experimental confirmation.\cite{Fradkin,Lee,Gorbar,Ludwig} Ludwig {\it et al.} obtained different
 minimum values, $\sigma_{\rm res}$, using two different
 approaches: $\sigma_{\rm res}=(\pi/2)e^2/h$ using the Kubo
 formula and $\sigma_{\rm res}=(4/\pi)e^2/h$ using a
 definition of conductivity in the sigma model.\cite{Ludwig} In many recent
 works, the issue has been further confused with findings of even more values of residual conductivity.
 In Refs.\,\onlinecite{Gusynin,Peres,Katsnelson,Tworzydlo} $\sigma_{\rm res}=(4/\pi)e^2/h$
 was obtained, while  Ziegler
predicted $\sigma_{\rm res}=\pi e^2/h$.\cite{Ziegler} $\sigma_{\rm
res}=(\pi/2) e^2/h$ was also obtained in
Ref.\,\onlinecite{Falkovsky}. More recently, Ziegler demonstrated
that all these values of $\sigma_{\rm res}$ can be obtained within
the Kubo formulism by taking different orders of the
zero-frequency dc and zero dissipation limits.\cite{Ziegler1}
Performing numerical calculations with the Kubo formula, Nomura
and MacDonald obtained $\sigma_{\rm res}=(4/\pi) e^2/h$ in the
case of short-range scattering and $\sigma_{\rm res}=4 e^2/h$ for
the Coulomb scattering case.\cite{MacDonald}

Employing the Boltzmann equation, Adam, {\it et al}. demonstrated
that the minimum "residual" conductivity arises from nonvanishing
electron density at zero gate voltage which may be induced by
impurity potentials.\cite{DasSarma} Analyzing random fluctuations
of gate voltage,  they found that the value of $\sigma_{\rm res}$
is not universal but depends on the impurity concentration. Apart
from this particular issue, the observed almost-linear variation
of conductivity with electron density can be easily understood
within the Kubo formula framework\cite{MacDonald} as well as with
the Boltzmann equation.\cite{DasSarma}

In this paper, we present a kinetic equation approach to investigate
transport in graphene considering long-range electron(or
hole)-impurity scatterings. Here, interband correlations
(polarization effects) are taken into account, whereas in all
previous studies, such interband correlations associated with
electron/hole-impurity scatterings were ignored. We find that the
conductivity in graphene contains two terms: one of which is
inversely proportional to impurity density, while the other one
varies linearly with the impurity density. This results in a minimum
(rather than "residual") conductivity at a nonvanishing critical
electron density, $N_c$, in the electron-density dependence of
conductivity. For electron density $N_{\rm e}$ larger than $N_c$,
the conductivity increases almost linearly with increasing $N_{\rm
e}$, while, the conductivity is approximately inversely proportional
to $N_{\rm e}$ for $N_{\rm e}<N_c$. We also demonstrate the effects
of various scattering potentials on the conductivity minimum.
Considering RPA screened electron-impurity scattering, we find that,
for typical experimental parameters, the critical electron density
is about $0.32N_i$ ($N_i$ is the impurity density) and the value of
minimum conductivity is equal to $5.1\,e^2/h$.

The paper is organized as follows. In Sec. II the kinetic equation
for nonequilibrium distribution functions as well as its solution
are presented. Also, the conductivity is exhibited in terms of
microscopically derived relaxation times. In Sec. III we present
our analytical results for the conductivity for several different
scattering potentials. Finally, the conclusions are summarized in
Sec. IV.

\section{Kinetic equation and solution}
\subsection{Kinetic equation}

In the Brillouin zone of graphene, there are six points at which
the energy of carriers vanishes and the conductance band touches
the valence band: ${\bf p}=(\pm 4\pi/(3\sqrt{3}a),0)$, ${\bf
p}=(2\pi/(3\sqrt{3}a),\pm 2\pi/(3a))$, and ${\bf
p}=(-2\pi/(3\sqrt{3}a),\pm 2\pi/(3a))$ with $a$ as the lattice
spacing. These points correspond to two inequivalent Dirac nodes,
$K$ and $K'$. In present paper, we are interested in the transport
of carriers in graphene with momenta near these Dirac points. The
Hamiltonian of an electron with two-dimensional momentum, ${\bf
p}\equiv (p_x,p_y)=(p\cos\phi_{\bf p},p\sin\phi_{\bf p})$, near
the $v=K$ or $K'$ Dirac nodes can be written as
\begin{equation}
\check h^{(v)}_0({\bf p})=\gamma [\hat\sigma_x p_x+{\rm
sgn}(v)\hat\sigma_y p_y],\label{H}
\end{equation}
with ${\rm sgn}(v)=1$ or $-1$ for $v=K$ or $K'$, and $\gamma\equiv
\sqrt{3}\alpha a/2$ is a material constant ($\alpha$ is the hopping
parameter in tight-binding approximation). $\gamma$ is equal to the
Fermi velocity, which is independent of carrier density in graphene.

The Hamiltonian (\ref{H}) can be diagonalized, resulting in two
eigen wavefunctions, $\varphi_{\mu{\bf p}}^{(v)}({\bf r})\equiv
u_\mu ({\bf p}){\rm e}^{i{\bf p}\cdot {\bf r}}$, and two
eigenvalues, $\varepsilon_\mu=(-1)^{\mu+1} e_{\bf p}$, with
$\mu=1,2$ as the helicity index, $e_{\bf p}\equiv \gamma p$, and
\begin{equation} u_\mu^{(v)} ({\bf p})
=\frac{1}{\sqrt{2}}\left (
\begin{array}{c}
{\rm e}^{(-1)^{\mu}i{\rm sgn}(v)\phi_{\bf p}}\\
1
\end{array}
\right ).
\end{equation}
$\varepsilon_1({\bf p})$ and $\varepsilon_2({\bf p})$ are just the
dispersion relations of the conduction and valence bands,
respectively.

It is useful to introduce a unitary transformation, $U_{\bf
p}=(u_1({\bf p}),u_2({\bf p}))$, which corresponds to a change from
a pseudospin basis to a pseudo-helicity basis. Applying this
transformation, the Hamiltonian (\ref{H}) is diagonalized as $\hat
h^{(v)}_0({\bf p})\equiv [U^{(v)}_{\bf p}]^+{\check h^{(v)}}_0({\bf
p})U^{(v)}_{\bf p} ={\rm diag}[\varepsilon_1({\bf
p}),\varepsilon_2({\bf p})]$. Note that $\hat h^{(v)}_0({\bf p})$ is
independent of node index $v$, indicating the existence of a valley
degeneracy.

In a realistic graphene system, the carriers experience scattering
by impurities. We assume that, in the pseudospin basis, the
interaction between carriers and impurities can be characterized by
an isotropic potential, $V (|{\bf p}-{\bf k}|)$, which corresponds
to scattering a carrier from state ${\bf p}$ to state ${\bf k}$. In
the pseudo-helicity basis, the scattering potential takes the
transformed form, $\hat T_v({\bf p},{\bf k}) = [U^{(v)}_{\bf p}]^+ V
(|{\bf p}-{\bf k}|)U^{(v)}_{\bf k}$.

We are interested in the current in a graphene system driven by a
dc electric field, ${\bf E}$. In the pseudo-spin basis, this
electric field can be described by a scalar potential, $V=e{\bf
E}\cdot {\bf r}$, with ${\bf r}$ as the carrier coordinate. From
Eq.\,(\ref{H}) it follows that the pseudo-spin single-particle
current operator, $\check {\bf j}^{(v)}({\bf p})$, has vanishing
diagonal elements:
\begin{equation}
\check {\bf j}^{(v)}({\bf p})={\bm \nabla}_{\bf p} \check
h^{(v)}_0({\bf p}).\label{J}
\end{equation}
The observed net current, given by ${\bf J} =eg_s\sum_{v,\bf p}
{\rm Tr}[\check {\bf j}^{(v)}({\bf p})\check \rho^{(v)}({\bf p})]$
with $\check\rho^{(v)}({\bf p})$ as the pseudospin-basis
distribution function and $g_s$ as the spin degeneracy of
graphene, can be also determined in pseudo-helicity basis via
\begin{equation}
{\bf J}=g_se\sum_{v,\bf p} {\rm Tr}[\hat {\bf j}^{(v)}({\bf
p})\hat\rho^{(v)}({\bf p})],\label{J1}
\end{equation}
with $\hat {\bf j}^{(v)}({\bf p})\equiv [U^{(v)}_ {\bf p}]^+\check
{\bf j}^{(v)}({\bf p})U^{(v)}_{\bf p}$ and $\hat {\rho}^{(v)}({\bf
p})\equiv [U^{(v)}_ {\bf p}]^+\check {\rho}^{(v)}({\bf
p})U^{(v)}_{\bf p}$ being the pseudo-helicity-basis single-particle
current operator and distribution function, respectively.
Explicitly, Eq.\,(\ref{J1}) can be rewritten as
\begin{eqnarray}
{\bf J}=g_s\gamma e\sum_{{\bf p},v} \frac{1}{p}\left \{{\bf
p}\left [[\hat
\rho^{(v)}]_{11}({\bf p})-[\hat \rho^{(v)}]_{22}({\bf p})\right ]\right .\nonumber\\
\left . +2{\rm sgn}(v)[{\bf p}\times {\bf n}]{\rm Im}\left [[\hat
\rho^{(v)}]_{12}({\bf p})\right ]\right \},\label{J2}
\end{eqnarray}
with $[\rho^{(v)}]_{\mu\nu}({\bf p})$ as the elements of the
distribution function. In the derivation of Eq.\,(\ref{J2}), the
Hermitian feature of the distribution function, i.e.
$\hat\rho^{(v)}({\bf p}) = [\hat\rho^{(v)}]^+({\bf p})$, has been
used. From Eq.\,(\ref{J2}) it is evident that contributions to
current arise not only from the diagonal elements of the
distribution function, but also involve its off-diagonal elements.

To carry out the calculation of current in graphene, it is necessary
to determine the carrier distribution function. In the pseudo-spin
basis, the kinetic equation for the distribution, $\check
\rho^{(v)}({\bf p})$, can be written as
\begin{eqnarray}
e{\bf E}\cdot \nabla_{\bf p} \check \rho^{(v)}({\bf p}) +i[\check
h^{(v)}_0({\bf p}),\check \rho^{(v)}({\bf p})]=-\check
I^{(v)},\label{KEE1}
\end{eqnarray}
with $\check I^{(v)}$ as the collision term. Applying the unitary
transformation $U^{(v)}_{\bf p}$, the kinetic equation for the
distribution in the pseudo-helicity basis, $\hat \rho^{(v)}({\bf
p})$, takes the form
\begin{eqnarray}
e{\bf E}\cdot \left \{\nabla_{\bf p} \hat \rho^{(v)}({\bf p})+\left
[\hat \rho({\bf p}), \nabla_{\bf p} \left (U^{(v)}_{\bf
p}\right )^+U^{(v)}_{\bf p}\right ] \right \}\nonumber\\
+i[\hat h^{(v)}_0({\bf p}),\hat \rho^{(v)}({\bf p})]=-\hat
I^{(v)},\label{KQQ}
\end{eqnarray}
with $\hat I^{(v)}=[U^{(v)}_ {\bf p}]^+\check I^{(v)}[U^{(v)}_ {\bf
p}]$.

It should be noted that the second term on the left-hand side of
Eq.\,(\ref{KQQ}) is associated with interband tunneling since the
off-diagonal elements of matrix $e{\bf E}\cdot{\bm \nabla}_{\bf p}
\left (U^{(v)}_{\bf p}\right )^+U^{(v)}_{\bf p}$ are just the
interband-tunneling matrix elements $<u_{\mu}^{(v)}({\bf p})|e{\bf
E}\cdot{\bf r}|u_{\bar\mu}^{(v)}({\bf p})>$ ($\bar \mu=3-\mu$)
with ${\bf r}$ as carrier coordinate.\cite{Inter2,Interband} This
term of Eq.\,(\ref{KQQ}) results in a component of $[\hat
\rho^{(v)}]_{12}({\bf p})$ which depends on the strength of
electric field via ${\rm e}^{-b/E}$ ($b$ is a real, positive
parameter), and is much less than unity in the linear response
regime. Thus, the interband tunneling term of Eq.\,(\ref{KQQ}) is
not actually linear (despite its formal appearance), and, in fact,
it can be ignored at low fields. Hence, Eq.\,(\ref{KQQ}) can be
rewritten as
\begin{eqnarray}
e{\bf E}\cdot \nabla_{\bf p} \hat \rho^{(v)}({\bf p}) +i[\hat
h^{(v)}_0({\bf p}),\hat \rho^{(v)}({\bf p})]=-\hat
I^{(v)},\label{KEE}
\end{eqnarray}
with $\hat I^{(v)}$ as the collision term given by
\begin{equation}
\hat I^{(v)}= {\hat \Sigma}^r_{v\bf p}{\hat {\rm G}}^<_{v\bf
p}+{\hat \Sigma}^<_{v\bf p}{\hat {\rm G}}^a_{v\bf p}- {\hat {\rm
G}}^r_{v\bf p} {\hat \Sigma}^<_{v\bf p}-{\hat {\rm G}}^<_{v\bf
p}{\hat \Sigma}^a_{v\bf p}.\label{CT}
\end{equation}
Here, ${\hat {\rm G}}^{r,a,<}_{v\bf p}$ and ${\hat
\Sigma}^{r,a,<}_{v\bf p}$, respectively, are the nonequilibrium
Green's functions and self-energies for carriers near node $v=K$
or $K'$.

 In
the kinetic equation above, electron-impurity scattering is embedded
in the self-energies, ${\check \Sigma}^{r,a,<}_{v\bf p}$. In the
present paper, we only consider electron-impurity collisions in the
self-consistent Born approximation. It is widely accepted that this
is sufficiently accurate to analyze transport properties in the
diffusive regime.\cite{Jauho} Accordingly, the self-energies take
the forms:
\begin{equation}
\hat{\Sigma}^{r,a,<}_{v\bf p}=n_i\sum_{{\bf k}}\hat T_v({\bf
p},{\bf k})\hat{\rm G}^{r,a,<}_{v\bf k}\hat T_v^+({\bf p},{\bf
k}).\label{SE}
\end{equation}

In the present paper, we restrict our considerations to the linear
response regime. In connection with this, all the functions, such
as the nonequilibrium Green's functions, self-energies, and
distribution function, can be expressed as sums of two terms: $A =
A_0 + A_1$, with $A$ representing the Green's functions,
self-energies or distribution function. $A_0$ and $A_1$,
respectively, are the unperturbed part and the linear electric
field part of $A$. In these terms, the kinetic equation for the
linear electric field part of the distribution, $\hat
\rho^{(v)}_1({\bf p})$, takes the form
\begin{eqnarray}
e{\bf E}\cdot \nabla_{\bf p}\hat \rho^{(v)}_0({\bf p})
 +i\left [\hat h^{(v)}_0,\hat
\rho^{(v)}_1({\bf p})\right ]=-\hat I_v^{(1)},\label{KE}
\end{eqnarray}
with $\hat I_v^{(1)}$ as the linear electric field part of the
collision term $\hat I_v$. This equation can be further rewritten
explicitly as
\begin{equation}
e{\bf E}\cdot \nabla_{\bf p}[\hat \rho^{(v)}_0]_{\mu\mu}({\bf
p})=-[\hat I^{(1)}_s]_{\mu\mu},\label{KED1}
\end{equation}
and
\begin{eqnarray}
2i\gamma p[\hat \rho^{(v)}_1]_{12}({\bf p})=-[\hat
I^{(1)}_s]_{12},\label{KEO1}
\end{eqnarray}
with $(\hat \rho_1)_{\mu\nu}({\bf p})$ and $(\hat
\rho_0)_{\mu\mu}({\bf p})$, respectively, as the elements of $\hat
\rho_1({\bf p})$ and  $\hat \rho^{(v)}_0({\bf p})={\rm
diag}[n_{\rm F}(\varepsilon_1({\bf p})),n_{\rm
F}(\varepsilon_1({\bf p}))]$, and ${\bf n}$ is a unit vector
perpendicular to the graphene plane.

To further simplify Eq.\,(\ref{KE}), we employ a two-band
generalized Kadanoff-Baym ansatz (GKBA).\cite{GKBA,GKBA1} This
ansatz, which expresses the lesser Green's function in terms of
the Wigner distribution function, has been proven sufficiently
accurate to analyze transport and optical properties in
semiconductors.\cite{Jauho} To first order in the dc field
strength, the GKBA reads,
\begin{equation}
\hat {\rm G}^<_{1}({\bf p},\omega)=-\hat {\rm G}_{0}^r({\bf
p},\omega)\hat \rho_1({\bf p})+\hat \rho_1({\bf p})\hat {\rm
G}_{0}^a({\bf p},\omega),\label{GKBA1}
\end{equation}
where the unperturbed retarded and advanced Green's functions are
diagonal: $\hat {\rm G}_0^{r,a}({\bf p},\omega)={\rm
diag}[(\omega-\varepsilon_1({\bf p})\pm
i\delta)^{-1},(\omega-\varepsilon_2({\bf p})\pm i\delta)^{-1}]$.
In our treatment, the effect of $\hat {\rm G}_1^{r,a}({\bf
p},\omega)$ on the distribution function has been ignored because
these linear electric field parts of the retarded and advanced
Green's functions lead to a collisional broadening effect on $\hat
\rho_1({\bf p})$, which plays a secondary role in transport
studies. Further, in our treatment, we ignore intervalley
transitions of carriers between different Dirac points since the
corresponding rates are very small in the presence of
carrier-impurity scattering.

Within the framework of these considerations, the scattering term
$\hat I^{(1)}_v$ in Eq.\,(\ref{KE}) can be expressed in terms of
the distribution function: its diagonal elements, $[\hat
I^{(1)}_v]_{\mu\mu}$, can be written as
\begin{eqnarray}
[\hat I^{(1)}_v]_{\mu\mu}=\pi N_i\sum_{{\bf k}}|V({\bf p}-{\bf
k})|^2\delta[\varepsilon^{(v)}_\mu({\bf
p})-\varepsilon^{(v)}_\mu({\bf
k})]\nonumber\\
\times \left \{[1+\cos(\phi_{\bf p}-\phi_{\bf k})]\{[\hat
\rho_1^{(v)}]_{\mu\mu}({\bf p})-[\hat
\rho_1^{(v)}]_{\mu\mu}({\bf k})\}\right .\nonumber\\
-(-1)^\mu {\rm sgn}(v)\sin(\phi_{\bf p}-\phi_{\bf k}) \left . {\rm
Im} \{[\hat \rho_1^{(v)}]_{12}({\bf p})+[\hat
\rho_1^{(v)}]_{12}({\bf k})\}\right \}
\end{eqnarray}
while the off-diagonal element, $[\hat I^{(1)}_v]_{12}$, takes the
form
\begin{eqnarray}
[\hat I^{(1)}_v]_{12}=\frac {\pi N_i}{2}\sum_{{\bf k},\mu}|V({\bf
p}-{\bf k})|^2\delta[\varepsilon^{(v)}_\mu({\bf
p})-\varepsilon^{(v)}_\mu({\bf
k})]\nonumber\\
\times \left \{(-1)^{\mu+1} i {\rm sgn}(v)\sin(\phi_{\bf
p}-\phi_{\bf k})\{[\hat \rho_1^{(v)}]_{\mu\mu}({\bf p})-[\hat
\rho_1^{(v)}]_{\mu\mu}({\bf k})\}\right .\nonumber\\
+[1- \cos(\phi_{\bf p}-\phi_{\bf k})] \left.\{[\hat
\rho_1^{(v)}]_{12}({\bf p})-[\hat \rho_1^{(v)}]_{21}({\bf
k})\}\right \}.
\end{eqnarray}
In the derivation of these equations, the effects of the real parts
of the retarded and advanced Green's functions on $\hat I^{(1)}_v$
have been ignored.

\subsection{Conductivity and solution of the kinetic equation}
Since $\varepsilon_\mu({\bf p})$, as well as the equilibrium
distribution $\hat \rho^{(v)}_0({\bf p})$, depend only on the
magnitude of momentum, the dependence of $\hat \rho^{(v)}_1({\bf
p})$ on momentum angle can be evaluated explicitly. From
Eq.\,(\ref{KE}) we see that the diagonal elements of $\hat
\rho^{(v)}_1({\bf p})$ can be written as [${\bf v}^{(v)}_\mu({\bf
p})={\bm \nabla}_{\bf p}\varepsilon_\mu^{(v)}({\bf
p})=(-1)^{\mu+1}\gamma {\bf p}/p$]
\begin{equation}
[\hat\rho_1^{(v)}]_{\mu\mu}({\bf p})=e{\bf E}\cdot {\bf
v}^{(v)}_\mu({\bf p}) \Lambda^{(v)}_\mu(p),\label{R1D}
\end{equation}
while the off-diagonal element, $[\hat\rho_1^{(v)}]_{12}({\bf p})$
, takes the form
\begin{equation}
[\hat\rho_1^{(v)}]_{12}({\bf p})=\frac{e\gamma}{p}[{\bf E}\times
{\bf p}\cdot {\bf n}] \Phi^{(v)}(p).\label{R1O}
\end{equation}
The functions $\Lambda^{(v)}_\mu(p)$ and $\Phi^{(v)}(p)$ depend
only on the magnitude of momentum and are determined by coupled
equations:
\begin{eqnarray}
-\frac{\partial \{[\hat \rho^{(v)}_0]_{\mu\mu}({\bf
p})\}}{\partial \varepsilon^{(v)}_\mu({\bf
p})}=\frac{1}{\tau_\mu^{(a)}(p)}\left \{\Lambda_\mu^{(v)}(p)-{\rm
sgn}(v){\rm Im}[\Phi^{(v)}(p)]\right \},\label{KEF1}
\end{eqnarray}
\begin{eqnarray}
2\gamma p{\rm
Re}[\Phi^{(v)}(p)]=\sum_\mu\frac{1}{2\tau^{(a)}_\mu(p)}\nonumber\\
\times\left \{{\rm sgn}(v)\Lambda_\mu^{(v)}(p)-{\rm
Im}[\Phi^{(v)}(p)]\right \},\label{KEF2}
\end{eqnarray}
and
\begin{eqnarray}
2\gamma p{\rm Im}[\Phi^{(v)}(p)]
=\sum_\mu\frac{1}{2\tau^{(b)}_\mu(p)}\left \{{\rm
Re}[\Phi^{(v)}(p)]\right \},\label{KEF3}
\end{eqnarray}
with $\tau^{(a,b)}_\mu(p)$ as microscopically determined
relaxation times independent of node index $v$:
$[\tau^{(a,b)}_\mu(p)]^{-1}=\pi N_i\sum_{\bf k}|V({\bf p}-{\bf
k})|^2\delta[\varepsilon_\mu({\bf p})-\varepsilon_\mu({\bf
k})]A^{(a,b)}(\phi_{\bf k})$; $A^{(a)}(\phi)=\sin^2\phi$, and
$A^{(b)}(\phi)=(1-\cos\phi)^2$. Solving the algebraic equations
(\ref{KEF1})-(\ref{KEF3}) yields explicit analytic expressions for
the functions $\Lambda_\mu^{(v)}$ and $\Phi^{(v)}(p)$ in terms of
the relaxation times as
\begin{eqnarray}
{\rm Re}[\Phi^{(v)}(p)]&=&-\frac{{\rm
sgn}(v)}{4\gamma^2p^2}\sum_\mu\left \{\gamma p\frac{\partial
\{[\hat \rho_0^{(v)}]_{\mu\mu}({\bf p})\}}{\partial
\varepsilon^{(v)}_\mu({\bf p})}\right \},
\end{eqnarray}
\begin{eqnarray}
{\rm
Im}[\Phi^{(v)}(p)]=-\frac{{\rm sgn}(v)}{16\gamma^3p^3}\left[\frac{1}{\tau^{(b)}_1(p)}+\frac{1}{\tau^{(b)}_2(p)}\right ]\nonumber\\
\times\sum_\mu\left [\gamma p\frac{\partial \{[\hat
\rho_0^{(v)}]_{\mu\mu}({\bf p})\}}{\partial
\varepsilon^{(v)}_\mu({\bf p})}\right ],
\end{eqnarray}
and
\begin{eqnarray}
\Lambda^{(v)}_\mu(p)=-\tau^{(a)}_\mu(p)\frac{\partial \{[\hat
\rho_0^{(v)}]_{\mu\mu}({\bf p})]}{\partial
\varepsilon^{(v)}_\mu({\bf p})}\nonumber\\
+{\rm sgn}(v){\rm Im}[\Phi^{(v)}(p)].\label{L}
\end{eqnarray}

Substituting Eqs.\,(\ref{R1D}) and (\ref{R1O}) into
Eq.\,(\ref{J2}), ${\bf J}$ can be further rewritten as
\begin{equation}
{\bf J}=\frac {e^2}2 g_s\gamma^2{\bf E}\sum_{{\bf p},v}
\{[\Lambda^{(v)}_1(p)+
\Lambda^{(v)}_2({p})]\nonumber\\
+2{\rm sgn}(v){\rm Im}[\Phi^{(v)}({p})]\}.\label{J3}
\end{equation}
Inserting the explicit forms of $\Lambda^{(v)}_\mu(p)$ and
$\Phi^{(v)}({p})$, we find that the conductivity, $\sigma$, takes
the form
\begin{eqnarray}
\sigma=-\frac {e^2}2 g_s\gamma^2\sum_{{\bf p},\mu,v}\left \{\left
[\tau^{(a)}_\mu(p)+\frac{1}{4\gamma^2p^2\tau^{(b)}(p)}
\right ]\right .\nonumber\\
\times\left .\frac{\partial \left \{[\hat
\rho_0^{(v)}]_{\mu\mu}({\bf p})\right \}}{\partial
\varepsilon^{(v)}_\mu({\bf p})} \right
\},\,\,\,\,\,\,\,\label{sigma}
\end{eqnarray}
with $[\tau^{(b)}(p)]^{-1}\equiv
[\tau^{(b)}_1(p)]^{-1}+[\tau^{(b)}_2(p)]^{-1}$.

From Eq.\,(\ref{sigma}) it is clear that, the conductivity in
graphene involves not only a term proportional $N_i^{-1}$, but also
a term linear in impurity density. However, there is {\it no} term
independent of impurity scattering. This differs significantly from
previous results. Adams {\it et al.} and Galitski {\it et al.} found
the conductivity to always be proportional to
$N_i^{-1}$,\cite{DasSarma,DasSarma2} while the other previous
calculations indicated that the conductivity always contains a term
independent of impurity scattering.\cite{Ludwig,Gusynin,Peres,
Katsnelson,Tworzydlo,Ziegler,Falkovsky,Ziegler1,MacDonald}

 It should be
noted that the anomalous conductivity term proportional to $N_i$
arises from the nonvanishing of $[\hat\rho_1^{(v)}]_{12}({\bf p})$.
This can be seen from Eq.\,(\ref{J3}): the conductivity depends on
the imaginary part of $[\hat\rho_1^{(v)}]_{12}({\bf p})$, which is
proportional to $[{\tau^{(b)}_\mu(p)}]^{-1}$. Since
$[\hat\rho_1^{(v)}]_{12}({\bf p})$ describes interband coherence,
the anomalous conductivity term is a result of interband
correlation.

From Eqs.\,(\ref{KEF2}) and (\ref{KEF3}) it follows that the
imaginary part of $[\hat\rho_1^{(v)}]_{12}({\bf p})$ is proportional
to $N_i$, while its real part is independent of impurity density.
This can be understood from the fact that both the driving electric
field and impurity scattering can cause transitions between two
pseudo-helicity states: they result in a change of carrier momentum
but retain the pseudospin unchanged, leading to a change of
pseudo-helicity. Obviously, the probability of this transition  is
proportional to the strength of the momentum change, which is
determined by the electric field and/or by impurity scattering.
Thus, impurity scattering results in a term of
$[\hat\rho_1^{(v)}]_{12}({\bf p})$ proportional to $N_i$, while the
$N_i$-independent term of $[\hat\rho_1^{(v)}]_{12}({\bf p})$ is a
result of the momentum  change induced directly by the electric
field.

\section{Results and discussion}

As pointed out by Adam, {\it et al.}, "intrinsic graphene" with
chemical potential (or Fermi energy) precisely at the Dirac point
is experimentally unrealizable since charged impurity disorder or
spatial inhomogeneity in the system can lead to a nonvanishing
induced graphene carrier density.\cite{DasSarma} Accordingly, all
experimental graphene samples are extrinsic since some free
carriers always exist in the system. In connection with this, the
conductivity of graphene arises mainly from the contribution of
one type of carrier, electrons or holes, depending on the
experimental conditions.

In the present paper, we focus on electron transport in graphene
at zero temperature. In this case, the chemical potential (or,
equivalently the Fermi energy, $E_F$) is given by $E_F=\gamma k_F$
with the 2D Fermi wavevector $k_F$ depending on the electron
density through $k_F=\sqrt{4\pi N_{\rm e}/g_sg_v}$ ( we choose the
spin and valley degeneracies, respectively, as $g_s=g_v=2$ in the
present paper). The zero-temperature conductivity is given by
\begin{eqnarray}
\left .\sigma \right |_{T=0}={e^2} \frac{ g_s g_v}{4\pi}\gamma k_F \left
[\tau^{(a)}_1(k_F)+\frac{1}{4\gamma^2k_F^2\tau^{(b)}(k_F)}
\right ].\label{sigma_T=0}
\end{eqnarray}
To analyze the effects of scattering potentials on $\sigma$, we
consider short-range (SR), Thomas-Fermi (TF), as well as
random-phase-approximation (RPA) screened Coulomb interactions.

{\it Short-range screened Coulomb interaction-} For a short-range
screened interaction potential, $V({\bf p}-{\bf k})$ becomes
independent of electron momentum: $V({\bf p}-{\bf k})\approx
\pi\gamma/(2k_F)$.\cite{MacDonald} In connection with this, the
relaxation times $\tau^{(a)}_{1}(k_F)$ and $\tau^{(b)}(k_F)$,
respectively, reduce to $\tau^{(a)}_{1}(k_F)=16 N_{\rm e}/(N_i\pi
\gamma k_F)$ and
 $\tau^{(b)}(k_F)=8 N_{\rm e}/(3\pi N_i\gamma k_F)$. Thus, the
zero-temperature conductivity for a short-range potential, $\left
.\sigma \right |^{\rm SR}_{T=0}$, reduces to
\begin{equation}
\left .\sigma \right |^{\rm SR}_{T=0}= \frac{e^2}{\pi} \left
[\frac{16N_{\rm e}}{\pi N_i}+\frac{3\pi N_{ i}}{32 N_{\rm
e}}\right ].\label{final}
\end{equation}

\begin{figure}
\includegraphics [width=0.45\textwidth,clip] {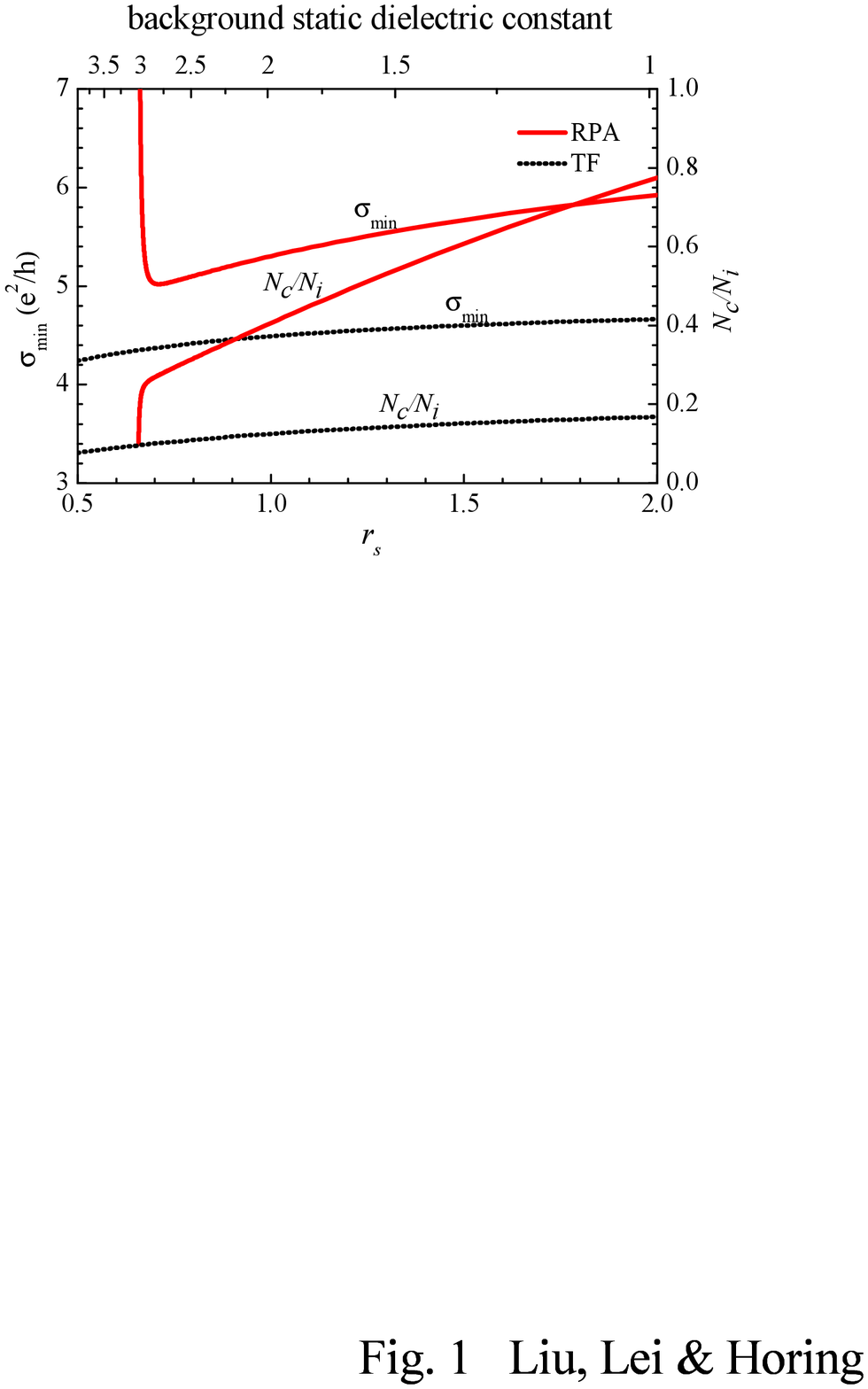}
\caption{The dependencies of minimum conductivity and critical
density on $r_s$ for TF and RPA screened Coulomb potentials.
$\gamma$ is chosen to be $1.1\times 10^5$m/s.} \label{fig1}
\end{figure}

{\it Thomas-Fermi screened potential-} We also evaluate the
conductivity for a 2D Thomas-Fermi screened potential, $V^{\rm
TF}(q)$, given by
\begin{equation}
V^{\rm TF}(q)=\frac{e^2}{2\pi\varepsilon_0 \kappa (q+q_s)},
\end{equation}
with ${\bf q}\equiv {\bf p}-{\bf k}$, $q_s=4e^2k_F/(\kappa
\gamma)$ and $\kappa$ as the static background dielectric
constant. In this case, the scattering times $\tau^{(a)}_{1}(k_F)$
and $\tau^{(b)}(k_F)$ can be obtained analytically as:
\begin{equation}
\left[ \frac{1}{\tau^{(a)}_{1}(k_F)}\right ]_{\rm
TF}=\frac{N_i}{N_{\rm e}}\gamma k_F G(2r_s),
\end{equation}
and
\begin{equation}
\left[ \frac{1}{\tau^{(b)}(k_F)}\right ]_{\rm
TF}=\frac{N_i}{N_{\rm e}}\gamma k_F F(2r_s),
\end{equation}
with $r_s=e^2/(4\pi\varepsilon_0\kappa\gamma)$. The functions
$G(x)$ and $F(x)$ are defined as
\begin{widetext}
\begin{eqnarray}
G(x)&=&\frac{x^2}{8}\int_0^{2\pi}d\theta \frac{\sin^2\theta}{(\sin\frac{\theta}{2}+x)^2} \nonumber\\
&=&x^2\left \{\begin{array}{ll}
 \frac{\pi}{4}+3x-\frac{3\pi}{2}x^2+{|x|(3x^2-2)\arccos(1/x)}[{x^2-1}]^{-1/2}
&|x|>1\\
 \frac{\pi}{4}+3x-\frac{3\pi}{2}x^2+{x(3x^2-2){\rm Re}\left [{\rm
arctanh}(1/\sqrt{1-x^2})\right ]}[{1-x^2}]^{-1/2} & 0\le x\le 1
\end{array}
\right .,
\end{eqnarray}
and
\begin{eqnarray}
F(x)&=&\frac{x^2}{8}\int_0^{2\pi}d\theta \frac{(1-\cos\theta)^2}{(\sin\frac{\theta}{2}+x)^2}\nonumber\\
&=&x^2\left \{\begin{array}{ll} \frac\pi
2-4x+3x^2\pi-\frac{2x^3}{x^2-1}+{|x^3|(8-6x^2)\arccos(1/x)}{(x^2-1)^{-3/2}}
& |x|>1\\
\frac\pi 2-4x+3x^2\pi-\frac{2x^3}{x^2-1}-{x^3(8-6x^2){\rm Re}\left
[{\rm arctanh}(1/\sqrt{1-x^2})\right ]}{(1-x^2)^{-3/2}} & 0\le
x\le 1
\end{array} \right . .
\end{eqnarray}
\end{widetext}
Thus, the conductivity for the TF potential, $\left .\sigma \right
|^{\rm TF}_{T=0}$, is determined by
\begin{equation}
\left .\sigma \right |^{\rm TF}_{T=0}= \frac{e^2}{\pi} \left
[\frac{N_{\rm e}}{ N_i G(2r_s)}+\frac{N_{i}F(2r_s)}{4 N_{\rm
e}}\right ].\label{final1}
\end{equation}

{\it RPA-screened Coulomb potential-} We also examine the
conductivity of graphene in the presence of RPA-screened
electron-impurity scattering, with the potential taking the form:
\begin{equation}
V(q)=\frac{e^2}{2\pi\varepsilon_0\kappa q\varepsilon(q)},
\end{equation}
and the dielectric function $\varepsilon(q)$ for the massless
Dirac energy spectrum is given by\cite{Ando,DasSarma1}
\begin{eqnarray}
\varepsilon(q)=1+\frac{q_s}{q}\left\{\theta(2k_F-q)\left
(1-\frac{\pi
q}{8 k_F}\right )\right.\nonumber\\
\left .+\theta(q-2k_F)\left [1-\frac 12 \sqrt{1-\left(2k_F/q\right
)^2}\right.\right .\nonumber\\
\left .\left . -\frac{q}{4k_F}\arcsin(2k_F/q)\right ]
\right\}.\label{DF}
\end{eqnarray}
Substituting the RPA-screened potential into the expressions for
the scattering times, we obtain
\begin{equation}
\left[ \frac{1}{\tau^{(a)}_{1}(k_F)}\right ]_{\rm
RPA}=\frac{N_i}{N_{\rm e}}\gamma k_F G[4r_s/(2-\pi r_s)],
\end{equation}
and
\begin{equation}
\left[ \frac{1}{\tau^{(b)}(k_F)}\right ]_{\rm
RPA}=\frac{N_i}{N_{\rm e}}\gamma k_F F[4r_s/(2-\pi r_s)].
\end{equation}
Hence, from Eq.\,(\ref{sigma_T=0}), it follows that the
conductivity for the RPA-screened Coulomb potential case is given
by
\begin{eqnarray}
\left .\sigma \right |^{\rm RPA}_{T=0}= \frac{e^2}{\pi} \left
[\frac{N_{\rm e}}{ N_i G[4r_s/(2-\pi r_s)]}\right .
\nonumber\\
\left . +\frac{N_{i}F[4r_s/(2-\pi r_s)]}{4 N_{\rm e}}\right
].\label{final2}
\end{eqnarray}

From Eqs.(\ref{final}), (\ref{final1}), and(\ref{final2}) it is
evident that a minimum exists in the electron-density dependence
of graphene conductivity for short-range, Thomas-Fermi as well as
RPA-screened Coulomb potentials. The corresponding critical values
of electron density, $N_c$, and the minimum conductivities,
$\sigma_{\rm min}$, are shown in table I. Furthermore, we verify
that, for $N_{\rm e}>N_c$, the conductivity varies almost linearly
with the electron density, while it is inversely proportional to
$N_{\rm e}$ when $N_{\rm e}<N_c$.

\begin{table*}[t]
\begin{center}
\begin{tabular*}{0.75\textwidth}{lll}
\hline
 &$N_c/N_i$&$\sigma_{\rm min}/(e^2/h)$\\
\hline SR&$\pi\sqrt{6}/32\approx 0.24$ &$2\sqrt{6}\approx 4.9$
\\

TF&$\sqrt{F(2r_s)G(2r_s)}/2$&$2[\sqrt{F(2r_s)/G(2r_s)}]$\\
RPA&$\sqrt{F[4r_s/(2-\pi r_s)]G[4r_s/(2-\pi r_s)]}/2$&$ 2
\{\sqrt{F[4r_s/(2-\pi r_s)]/G[4r_s/(2-\pi r_s)]}\}$ \\
\hline
\end{tabular*}
 \caption{Critical electron densities and minimum conductivities for the various screened
 Coulomb potentials considered}
 \end{center}
  \end{table*}

From Eq.\,(\ref{final}) we see that, for SR screened scattering,
$N_c/N_i$ and $\sigma_{\rm min}$ are constants independent of
$r_s$. However, for the TF and RPA potentials, they are functions
of $r_s$, which is, in principle, a tunable parameter through its
dependence on the background dielectric constant $\kappa$. In
Fig.\,1, we plot the dependencies of $N_c$ and $\sigma_{\rm min}$
on $r_s$. This figure enables us to determine the values of
$\sigma_{\rm min}$ and $N_c$ for various experimental samples.

Experimentally, graphene is usually fabricated using a SiO$_2$
substrate and hence the background static dielectric constant,,
$\kappa$, can be estimated as $\kappa=2.45$.\cite{DasSarma}
 The measured Fermi velocity $\gamma=1.1\times 10^5$m/s
 leads to $r_s\approx0.813$.\cite{Zhang} For this $r_s$ value, we obtain
\begin{eqnarray}
\sigma_{\rm min}^{\rm TF}\approx 4.42e^2/h,& N_c^{\rm TF}\approx 0.11N_i,\\
\sigma ^{\rm RPA}_{\rm min}\approx 5.11e^2/h,&N_c^{\rm RPA}\approx
0.32 N_i.
\end{eqnarray}
Furthermore, the asymptotic dependence of conductivity on electron
density as $N_{\rm e}$ moves away from the critical value is given
in the TF and RPA cases as:
\begin{equation}
\left .\sigma\right |^{\rm TF}_{T=0}\longrightarrow
\frac{e^2}{h}\left \{
\begin{array}{cc}
{20.0N_{\rm e}}/{N_i}\,\,\,\,\,\,&N_{\rm e}>N_c^{\rm TF}\\
{0.25N_{i}}/{N_{\rm e}}\,\,\,\,\,\,&N_{\rm e}<N_c^{\rm TF}\\
\end{array}
\right .,
\end{equation}

\begin{equation}
\left .\sigma\right |^{\rm RPA}_{T=0}\longrightarrow
\frac{e^2}{h}\left \{
\begin{array}{cc}
{8.0N_{\rm e}}/{N_i}\,\,\,\,\,\,&N_{\rm e}>N_c^{\rm RPA}\\
{0.8N_{i}}/{N_{\rm e}}\,\,\,\,\,\,&N_{\rm e}<N_c^{\rm RPA}\\
\end{array}
\right ..
\end{equation}

The experimentally observed residual conductivity is about $4
e^2/h$ and our obtained minimum conductivity is in good
quantitative agreement with it. The contention of Adam, {\it et
al}., about finite electron density at zero gate voltage is
plausible in describing the observed plateau near zero gate
voltage.\cite{DasSarma} However, the vagaries of earlier theories
described in the Introduction, and our own prediction of an
unlimited increase of conductivity with decreasing $N_{\rm e}$ for
$N_{\rm e}<N_c$ (which tends to infinity as $N_{\rm e}\rightarrow
0$) calls for a re-evaluation of the assumption of linearity. We
believe that the problematic theoretical density dependence of
conductivity for low $N_{\rm e}$ is associated with the fact that,
in the dilute limit, the response of the system caused by the
electric field can {\it not} be treated as linear in $|{\bf E}|$.
It is well known that linear response theory is valid only for
$e|{\bf E}|\alpha<\gamma k_F$ with $\alpha \equiv<{\bf r}>\approx
\gamma \tau$ as the average displacement induced by the electric
field ($\tau$ is the relaxation time). In the dilute limit, $k_F$
is very small and the requirement $e|{\bf E}|\alpha<\gamma k_F$ is
difficult to satisfy experimentally. Hence, our theory, as well as
other linear response theories, can not be employed to describe
the transport behavior of carriers in real systems with $N_{\rm
e}\rightarrow 0$.

\section{Conclusions}
We have formulated a kinetic equation to investigate dc transport
in graphene considering long-range electron-impurity scatterings.
In this study, we included the role of interband correlations. We
found that the conductivity contains two terms: one term is
inversely proportional to the impurity density while the other one
is linear in $N_i$. This results in a minimum in the density
dependence of conductivity for $N_{\rm e}$ equal to a critical
density, $N_c$. For $N_{\rm e}>N_c$ the conductivity varies almost
linearly with the electron density while, for $N_{\rm e}<N_c$,
$\sigma$ is approximately inversely proportional to $N_{\rm e}$.
We also discussed the effects of various scattering potentials on
the conductivity minimum. Using typical experimental parameters,
we found that for the RPA-screened Coulomb potential, $\sigma_{\rm
min}\approx 5.1\,e^2/h$ and $N_{c}\approx 0.32N_i$.

\begin{acknowledgments}
This work was supported by the Youth Scientific Research Startup
Funds of SJTU and by projects of the National Science Foundation
of China and the Shanghai Municipal Commission of Science and
Technology.
\end{acknowledgments}

\end{document}